\documentclass[a4paper,11pt]{article}

\usepackage[english]{babel}
\usepackage[a4paper]{geometry}
\usepackage{enumerate}  
\usepackage{amsmath}
\usepackage{amsthm}
\usepackage{amssymb}
\usepackage{hyperref}   
\usepackage{amssymb}
\usepackage{color}
\usepackage{graphicx}
\usepackage{bbm}

\def\bR{\mathbb{R}}
\def\bN{\mathbb{N}}

\def\cU{\mathcal{U}}
\def\cW{\mathcal{W}}
\def\cV{\mathcal{V}}
\def\cF{\mathcal{F}}
\def\cG{\mathcal{G}}
\def\cQ{\mathcal{Q}}

\def\cJ{\mathcal{J}}
\def\cC{\mathcal{C}}
\def\cL{\mathcal{L}}
\def\cN{\mathcal{N}}
\def\cE{\mathcal{E}}
\def\cK{\mathcal{K}}
\def\cH{\mathcal{H}}

\newcommand{\unit}{1\!\!1}

\newcommand{\exps}{\tau}       %small exponent cut-off N-dependent chi_L
\newcommand{\exph}{\varepsilon}  %high exponent cut-off N-dependent chi_H

\def\eps{\varepsilon}
\def\ph{\varphi}

\def\wt{\widetilde}
\def\wh{\widehat}

\def \pn {\varphi_0 }

\def \an {\mathfrak{a}_{0} }

\def \cFpn {\mathcal{F}_{\bot \varphi_0}^{\leq N} }
\def \wt {\widetilde}

\def\tr{\operatorname{tr }}
\def\be{\begin{equation}}
\def\ee{\end{equation}}

\newtheorem{theorem}{Theorem}[section]  % use thm for %Theorems to keep numbering consistent

\newtheorem{prop}[theorem]{Proposition}

\numberwithin{equation}{section}

\begin{document}

\title{The Low Energy Spectrum of Trapped Bosons \\ in the Gross-Pitaevskii Regime}

\author{Christian Brennecke   \\
\\
Institute for Applied Mathematics, University of Bonn, \\
Endenicher Allee 60, 53115 Bonn, Germany  \\
\\}

\maketitle

\begin{abstract}
Bogoliubov Theory \cite{Bog} provides important predictions for the low energy \linebreak properties of the weakly interacting Bose gas. Recently, Bogoliubov's predictions could be justified rigorously in \cite{BBCS3} for translation invariant systems in the Gross-Pitaveskii regime, where $N$ bosons in $\Lambda = [0;1]^3\subset \bR^3$ interact through a potential whose scattering length is of size $ N^{-1}$. In this note, we review recent results from \cite{BSS2}, a joint work with B. Schlein and S. Schraven, which extends the analysis of \cite{BBCS3} to systems of bosons  in $\bR^3$ that are trapped by an external potential. 
\end{abstract}

\section{Introduction and Main Results}  

The rigorous understanding of low energy properties of dilute Bose gases has been an active field of research in the last decades after the first experimental realization of Bose-Einstein condensates in trapped atomic gases \cite{CW95, K95}. As a mathematical model for such experimental setups, we consider systems of $N$ trapped bosons moving in $\bR^3$ in the Gross-Pitaevskii regime. %(see e.g. \cite[Chapters 5, 11, 12]{PS} and the references therein for relevant physics background). 
The energy of the system is described by 
\begin{equation}\label{eq:defHN} H_N = \sum_{j=1}^N \left( -\Delta_{x_j} + V_\text{ext} (x_j) \right) + \sum_{1\leq i<j\leq N} N^2 V(N(x_i -x_j)),
\end{equation}
acting in a dense subspace of $L^2_s (\bR^{3N})$, the subspace of $L^2 (\bR^{3N})$-functions which are invariant under permutations of the $N$ particle coordinates. We assume that the trapping potential $V_\text{ext}\in L^\infty_{loc}(\mathbb{R}^3)$ is such that $ V_\text{ext}(x)\to \infty $ for $|x|\to \infty$ and that $V\in L^3(\bR^3)$ is pointwise non-negative, radial and of compact support. 

We are interested in the computation of the low energy spectrum of $H_N$ in the limit $N\to \infty$. Notice that, due to the presence of a trapping potential, the particles typically move in a region of volume $\mathcal{O}(1)$. By a simple scaling argument, the system is equivalent to a system of $N$ particles interacting through the unscaled interaction $V$ and trapped in a region of volume $\mathcal{O}(N^3)$. That is, the Gross-Pitaevskii limit can be thought of as a joint thermodynamic and low density limit (with the particle density of size $N^{-2}$). In particular, in view of Bogoliubov theory \cite{Bog}, one may expect that the low energy spectrum depends on the interaction to leading order only through its scattering length.

Recall that the scattering length $\an$ of $V$ is defined through the solution $f$ of the zero energy scattering equation 
		\begin{equation}\label{eq:0en} \Big(- \Delta + \frac{1}{2} V \Big) f  = 0 \hspace{0.5cm} \text{with} \hspace{0.5cm} \lim_{|x|\to\infty} f(x) = 1. \end{equation}
For $ x \in \bR^3$ outside the support of $V$, $f$ is explicitly given by
		\[ f(x) = 1- \frac{\frak{a}_0 }{|x|} \]
for some constant $\frak{a}_0$, the \emph{scattering length of $V$}. A simple computation shows that 
		\begin{align*}  \frak{a}_0  = \frac1{8 \pi } \int_{\mathbb{R}^3} dx\,V(x) f(x)  \end{align*}
and that the solution of \eqref{eq:0en} with the rescaled potential $N^2V(N.)$ is equal to $f(N.)$, i.e.
		\[ \Big( - \Delta + \frac{1}{2} N^2 V (N.) \Big) f (N.) = 0. \]
This implies that the scattering length of the interaction $N^2 V (N.)$ is equal to $\frak{a}_0 N^{-1}$. The size $ N^{-1}$ of the scattering length characterizes the Gross-Pitaeavskii regime.  

The leading order contribution to the ground state energy $E_N$ of $H_N$ is well-known and has been derived in \cite{LSY, LS2,NRS}, showing that		
		\[\lim_{N\to \infty} \frac{E_N}N =  \inf_{\psi\in L^2(\bR^3): \|\psi\|_2=1}\cE_{GP}(\psi). 
		\]
Here, $\cE_{GP}$ denotes the Gross-Pitaevskii functional which is defined by
		\[
		\mathcal{E}_{GP}(\psi) = \int_{\bR^3} \left(  \vert \nabla \psi(x) \vert^2 + V_\text{ext}(x) \vert \psi(x)\vert^2 + 4\pi\an \vert \psi(x)\vert^4 \right) dx.
		\]
By standard arguments, $\cE_{GP}$ admits a unique normalized, strictly positive minimizer which we denote from now on by $\pn \in L^2(\bR^3)$. It solves the Euler-Lagrange equation 
\begin{align*} -\Delta \pn + V_\text{ext} \pn + 8\pi \frak{a}_0 |\pn|^2 \pn = \eps_{GP} \pn, \hspace{0.5cm} \text{where}\hspace{0.5cm} \eps_{GP} = \cE_{GP} (\pn) + 4\pi \frak{a}_0 \| \pn \|_4^4. \end{align*} 
%In the Appendix, more precisely in Lemma \ref{thm:gpmin1} and Lemma \ref{thm:gpmin2}, we establish regularity and decay properties of the minimizer $\pn$, which will be useful for our analysis. 

In order to determine the next to leading order contribution to the ground state energy as well as the excitation energies of $H_N$, we need to use the fact that approximate ground states exhibit complete Bose-Einstein condensation (BEC). Indeed, it follows from \cite{LS1, LS2, NRS} that every sequence of normalized wave functions $\psi_N \in L^2_s (\bR^{3N})$ with
		\[  \lim_{N \to \infty} \frac{1}{N} \langle \psi_N, H_N \psi_N \rangle = \cE_\text{GP} (\pn) \, ,  \]
that is, every sequence of approximate ground states, exhibits complete Bose-Einstein condensation into the minimizer $\pn$ of $ \cE_{GP}$. Mathematically, this means that the one-particle reduced density $\gamma_N^{(1)} =  \operatorname{tr}_{2,\dots,N} |\psi_N\rangle\langle\psi_N|$ associated with $\psi_N$ satisfies
		\begin{equation}\label{eq:BEC0}\lim_{N\to\infty} \langle\pn, \gamma_N^{(1)}\pn\rangle = 1. \end{equation}
Notice that this is equivalent to $ \lim_{N\to \infty} \gamma_N^{(1)}  = |\pn\rangle\langle \pn |$ in the trace class topology. 

In the sequel, we make use of a quantitative version of \eqref{eq:BEC0} with optimal rate of convergence that follows from \cite{BSS1}. More precisely, the main result of \cite{BSS1} is that (under the assumptions $(1)$ and $(2)$ in \eqref{eq:asmptsVVext} stated below) there exists some $C>0$ such that 
		\begin{equation}\label{eq:HNlwrbnd} H_N \geq N\mathcal{E}_\text{GP}(\pn) + C^{-1}\sum_{j=1}^N \big( 1- |\pn\rangle\langle\pn|\big)_{x_j}- C. \end{equation}
In particular, if $\psi_N\in L^2_s(\bR^{3N})$ is a normalized sequence of approximate ground states such that $ \langle \psi_N, H_N\psi_N\rangle \leq N\cE_\text{GP}(\pn) + \zeta $ for some $\zeta > 0$, then $\gamma_N^{(1)}$ satisfies 
		\[ 1-\langle\pn, \gamma_N^{(1)}\pn\rangle \leq C \frac{(C+\zeta)}N.\]
The lower bound \eqref{eq:HNlwrbnd} generalizes the main results of \cite{BBCS1, BBCS4}, dealing with translation invariant systems, to systems of trapped particles in $\bR^3$. For sufficiently small $\mathfrak{a}_0$, \eqref{eq:HNlwrbnd} has previously been proved in \cite{NNRT} (adapting a complete-the-square argument of \cite{BFS} to systems of trapped bosons) and in \cite{H} (for translation invariant systems, simplifying the arguments of \cite{BBCS1}). The bound \eqref{eq:HNlwrbnd} also follows from \cite{NT} (with no additional smallness restriction on the size of $\mathfrak{a}_0$) on which we comment briefly below. For further recent results on BEC, including regimes beyond the Gross-Pitaevskii scaling, see \cite{ABS, F}.
		
To state our main results, we denote by $H_{GP}$ the self-adjoint operator
		\be \label{eq:defHGPE} H_\text{GP} = -\Delta + V_\text{ext} +8\pi \mathfrak{a}_0\pn^2 -\eps_{GP} \ee 
and we make our assumptions on $V$ and $V_{ext}$ more precise:
		\begin{equation}\label{eq:asmptsVVext}
		\begin{split}
		(1) \;& V\in L^3(\bR^3) , V(x)\geq 0 \text{ for } a.e. \;x\in\bR^3, V(x)=V(y) \text{ if } |x|=|y|,  \\ &V  \text{ compactly supported}, \\
		(2)\; &   V_\text{ext} \in C^1\big(\bR^3; [0;\infty)\big), V_\text{ext}(x) \to \infty \text{ as } |x| \to \infty,\\ %V_\text{ext} (x)=V_\text{ext}(y) \text{ if } |x|=|y|, \\
		& \exists\;  C > 0 \; \forall \;  x,y\in\bR^3: V_\text{ext}(x+y) \leq C (  V_\text{ext}(x)+C)(  V_\text{ext}(y)+C), \\
		& \nabla V_\text{ext} \text{ has at most exponential growth as }  |x|\to \infty, \\
		(3) \;&(H_\text{GP} + 1)^{-3/4-\eps} e^{-\alpha |x|} \text{ is a Hilbert-Schmidt operator, for all $\eps > 0$} \\ &\text{and some $\alpha > 0$}.
		\end{split}
		\end{equation}
%Assumption (2) implies that $V_\text{ext},  \nabla V_\text{ext} $ grow at most exponentially fast as $|x|\to \infty$. Combined with the exponential decay of $\pn$, we use this to control the $L^p(\bR^3)$ norms of $V_\text{ext}\pn, (\nabla V_\text{ext})\pn$. The positivity assumption $V_{ext}\geq 0$ is for computational convenience only (it can be arranged by an energy shift).  
%Both $H_\text{GP}$ and $E$ have discrete spectrum and $\pn$ is the unique, positive and normalized ground state vector with $H_\text{GP}\pn = E \pn=0$.

\begin{theorem}\label{thm:main} \emph{(\cite[Theorem 1.1]{BSS2})}. Let $V$, $ V_\text{ext}$ and $H_{GP}$ satisfy \eqref{eq:asmptsVVext}. Then, there exists $\rho>0$ so that, in the limit $N\to \infty$, the ground state energy $E_N$ of $H_N$ is given by
		\be \begin{split}\label{eq:gsenergy}
		E_N &= N  \cE_{GP}(\pn) - 4\pi \mathfrak{a}_0 \|\pn\|_4^4 +E_{\text{Bog}}   + \mathcal{O}\big( N^{-\rho}\big),   
		\end{split}\ee
where $E_\text{Bog}$ is the finite constant (independent of $N$)  
		\begin{equation}\label{eq:bog1} 	
		\begin{split}
		E_\text{Bog} = \frac{1}{2} \lim_{\delta \to 0} &\left\{ \tr_{\perp \pn} \Big[ \Big(H_\text{GP}^{1/2} (H_\text{GP} + 16 \pi \frak{a}_0 \unit_\delta \pn^2 ) H_\text{GP}^{1/2} \Big)^{1/2} - H_\text{GP} - 8\pi \frak{a}_0 \unit_\delta \pn^2 \Big] \right. \\ &\hspace{5.5cm}  \left. + \frac{(8\pi\frak{a}_0)^2}{2} \tr  \Big[ \pn^2 \unit_\delta \frac{1}{-\Delta} \unit_\delta \pn^2 \Big] \right\} \end{split}  
		\end{equation}  
with $\unit_\delta$ denoting the approximation of the identity with $\unit_\delta (x;y) = (2\pi \delta)^{-3/2} e^{-(x-y)^2/2\delta^2}$ and where $\tr_{\bot \pn}$ denotes the trace over the orthogonal complement $\{\pn\}^{\bot}$. 
	
Moreover, the spectrum of $H_N-E_N$ below a threshold $\zeta>0$ (assuming $\zeta \leq CN^{\rho/5-\eps}$ for some sufficiently small $\eps>0$) consists of finite sums of the form 
	\be \label{eq:excitHN-EN}
	\sum_{i = 1}^\infty n_i e_i + \mathcal{O}\big( N^{-\rho} ( 1+ \zeta^5 + N^{-5\rho} \zeta^{30} )\big),
	\ee
	where $(e_j)_{j\in \mathbb{N}}$ denote the eigenvalues of the operator $E$ which is defined by
	\be \label{eq:defHGPE2}   E = \Big(H_\text{GP}^{1/2} \big(H_\text{GP}+ 16\pi \mathfrak{a}_0\pn^2\big)H_\text{GP}^{1/2} \Big)^{1/2}\ee
	and where $n_i \in \mathbb{N}$ with $n_i \neq 0$ for finitely many $i\in \mathbb{N}$.
\end{theorem}

\vspace{0.2cm}

\noindent {\textbf {Remarks:}} 
\begin{enumerate}[1)]

\item The form \eqref{eq:excitHN-EN} of the excitation spectrum of $H_N$ was conjectured in \cite{GS}, providing results comparable to Theorem \ref{thm:main} in the mean field limit (see also \cite{Sei,DN,LNSS} for previous and related results on the derivation of the excitation spectrum for mean field systems of bosons). In the mean field regime, more precise expansions in powers of $N^{-1}$ for the ground state energy and low energy excitation energies have been obtained in \cite{P3,BPS}. For further background and results on mean field systems, see for instance the recent survey article \cite{R}. 

\item Theorem \ref{thm:main} generalizes the main results of \cite{BBCS3} (dealing with translation invariant systems) to systems of trapped bosons. On a formal level, one can recover the results of \cite{BBCS3} by setting $ V_{ext}=0$ in $\Lambda = [0;1]^3$,  $V_{ext}=\infty$ otherwise and by imposing periodic boundary conditions in $\Lambda $. In this setting, the condensate wave function is described by $\ph= 1_{|\Lambda} \in L^2(\Lambda)$ and $H_{GP} = -\Delta $. In particular, $H_{GP}$ commutes with the multiplication operator $ 16\pi \mathfrak{a}_0\pn^2=16\pi \mathfrak{a}_0$ and $E$, defined in \eqref{eq:defHGPE2}, is the self-adjoint operator that multiplies in Fourier space by $ \sqrt{ |p|^4 + 16\pi \mathfrak{a}_0|p|^2} $, for $ p\in 2\pi \mathbb{Z}^3$. For trapped systems as in \cite{BSS2}, on the other hand, $H_{GP}$ and $ 16\pi \mathfrak{a}_0\pn^2$ do not commute anymore which makes the analysis technically more involved, compared to \cite{BBCS3}.

\item In (\ref{eq:bog1}), we cannot take directly the limit $\delta \to 0$, replacing $\unit_\delta$ with the identity, because the resulting operator is not a trace class operator. However, using the assumptions \eqref{eq:asmptsVVext}, we can compute the limit defining $E_\text{Bog}$ explicitly and find that
		\be \label{eq:Ebog}
		\begin{split}
		E_{\emph{Bog}} & = \frac{\kappa^2}{4} (8\pi \mathfrak{a}_0)^2 \tr \left[    \pn^2  (-\Delta+\kappa^2)^{-1} (-\Delta)^{-1}  \pn^2 \right]\\
		&\hspace{0.5cm} + \frac14 (8\pi\mathfrak{a}_0)^2 \left\| \pn    (-\Delta+\kappa^2)^{-1}[\pn^2,-\Delta] (-\Delta+\kappa^2)^{-1/2}   \right\|_\text{HS}^2  \\
		&\hspace{0.5cm} + \frac14 (8\pi\mathfrak{a}_0)^2 \left\| \pn   (-\Delta+\kappa^2)^{-1} \nabla\pn   \right\|_\text{HS}^2 \\
		&\hspace{0.5cm} + \frac{(8\pi \mathfrak{a}_0)^2}{4} \tr \, \Big[ \pn^2 \Big( \frac{1}{-\Delta + \kappa^2} - \frac{1}{H_\text{GP} + \kappa^2} \Big) \pn^2 \Big] \\
		&\hspace{0.5cm} + \frac{(8\pi \mathfrak{a}_0)^2}{4\kappa^2} \| \pn^3 \|^2 + \frac{(8\pi \mathfrak{a}_0)^2}{4}  \Vert (H_\text{GP}+\kappa^2)^{-1/2} Q \pn^3\Vert^2 \\ 
		&\hspace{0.5cm}-\frac{(8\pi \mathfrak{a}_0)^2 \kappa^2}{4} \tr_{\perp \pn} \Big[ \pn^2 Q \frac{1}{H_\text{GP} (H_\text{GP} + \kappa^2)} Q \pn^2 \Big] \\
		&\hspace{0.5cm} - \frac{(8\pi\frak{a}_0)^2}{\pi} \int_0^\infty ds \, \sqrt{s} \, \tr_{\perp \pn}  \frac{H_\text{GP}^{1/2}}{s+ H_\text{GP}^2} \Big[ \Big[ \pn^2 , \frac{H_\text{GP}}{s+ H_\text{GP}^2} \Big] , \pn^2 \Big] \frac{H_\text{GP}^{1/2}}{s+ H_\text{GP}^2} \\
		&\hspace{0.5cm} - \frac{4(8\pi\frak{a}_0)^3}{\pi} \int_0^\infty ds \, \sqrt{s}  \tr_{\perp \pn} \Big[ \frac{1}{s+H_\text{GP}^2} H_\text{GP}^{1/2} \pn^2 H_\text{GP}^{1/2} \Big]^3 \\ &\hspace{6cm} \times \frac{1}{s+H_\text{GP}^{1/2} (H_\text{GP} + 16 \pi \frak{a}_0 \pn^2) H_\text{GP}^{1/2}} 
		\end{split} \ee
for any $\kappa > 0$ large enough. In particular, all contributions on the right hand side of (\ref{eq:Ebog}) are finite. 

\item The form of the excitation spectrum \eqref{eq:excitHN-EN} has been derived independently in \cite{NT}, valid under slightly more general assumptions on $V$ and $V_{ext}$, compared to \eqref{eq:asmptsVVext}. We refer to \cite{NT, BSS2} for detailed comparisons of the assumptions and the proofs.

\item The formulas \eqref{eq:gsenergy} and \eqref{eq:Ebog} show that $E_N$ depends on the interaction $V$ only through its scattering length $\an$, up to errors vanishing in the limit $N\to \infty$. In particular, \eqref{eq:gsenergy} is the analogue of the Lee-Huang-Yang formula \cite{HY, LHY,LY}
		\be\label{eq:LHY} E_{N,L} = 4\pi \rho \an \Big( 1 + \frac{128}{15\sqrt{\pi}}  ( \rho \an^3)^{1/2} + o\big( ( \rho \an^3)^{1/2} \big)  \Big) \ee			
for the ground state energy $E_{N,L}$ of $N$ bosons trapped in $\Lambda_L = [0;L]^3$, in the thermodynamic limit where $N, L\to \infty$ with the density $\rho= NL^{-3}$ fixed. \eqref{eq:LHY} has been rigorously established in \cite{YY, BCS} (upper bound) and \cite{FS1, FS2} (lower bound). 

\end{enumerate}

In the following sections, we outline the main steps that lead to the proof of Theorem \ref{thm:main}. Our proof is based on a rigorous implementation of Bogoliubov's method \cite{Bog},\linebreak previously established in \cite{BBCS3} for translation invariant systems in the Gross-Pitaevskii regime. For the details of the implementation, we refer to \cite{BSS2}.

\section{Excitation Hamiltonians} \label{sec:fock}
%%%%%%%%%%%%%%%%%%%%%%%%%%%%%%%%%%%%%%%%%%%%%%%%%%%%%%%%%%%%
%%%%%%%%%%%%%%%%%%%%%%%%%%%%%%%%%%%%%%%%%%%%%%%%%%%%%%%%%%%%
%%%%%%%%%%%%%%%%%%%%%%%%%%%%%%%%%%%%%%%%%%%%%%%%%%%%%%%%%%%%

For our analysis, it is convenient to switch to a Fock space setting, factoring out the Bose-Einstein condensate. This allows us to focus on the analysis of the orthogonal excitations around the condensate. We follow here \cite{LNSS} and consider the unitary map $U_N : L^2_s (\bR^{3N}) \to \cF_{\perp \pn}^{\leq N}$ defined by $U_N \psi_N = ( \alpha_0, \alpha_1, \dots , \alpha_N )$ if
		\[ \psi_N = \alpha_0 \pn^{\otimes N} + \alpha_1 \otimes_s \pn^{\otimes (N-1)} + \ldots + \alpha_N \]
with $\alpha_j \in L^2_{\perp \pn} (\bR^3)^{\otimes j}$, for $j=0,1,\dots , N$. Here,  $\otimes_s$ denotes the symmetric tensor product and $\cF_{\perp \pn}^{\leq N}$ denotes the truncated excitation Fock space
		\[ \cF_{\perp \pn}^{\leq N} = \bigoplus_{n = 0}^N L^2_{\perp \pn} (\bR^3)^{\otimes_s n} \]
that is built over the orthogonal complement $  L^2_{\perp \pn} (\bR^3) = \{\pn\}^\bot$. The property \eqref{eq:BEC0} of BEC translates in the Fock space setting to the statement that
		\[ \lim_{N\to\infty} \Big( 1 - \langle\pn, \gamma_N^{(1)}\pn\rangle  \Big) = \lim_{N\to\infty} N^{-1} \langle  U_N\psi_N, \cN U_N\psi_N\rangle =0, \]
where $\cN$ denotes the number of particles operator in $\cF_{\perp \pn}^{\leq N}$, defined by $ (\cN \xi)^{(n)} = n \xi^{(n)} $ for every $\xi = (\xi^{(1)}, \dots, \xi^{(N)})\in \cF_{\perp \pn}^{\leq N}$. In other words, condensation into $\pn$ means that the average number of excitations is small compared to $N$, the total number of particles.

In $\cF_{\perp \pn}^{\leq N}$, we then consider $ \cL_N = U_N H_N U_N^*$ which reads in second quantized form 
	\begin{equation}\label{eq:cLNj}
	\begin{split}
	 \mathcal{L}_N =\;& \big\langle \pn, \big( -\Delta + V_\text{ext} + \frac{1}{2} \big(N^3 V(N\cdot) * \vert \pn \vert^2\big) \big) \pn \big\rangle (N-\cN) \\
	& \qquad -  \frac12\big\langle \pn,  \big(N^3 V(N\cdot) * \vert \pn \vert^2 \big)\pn \big\rangle (\cN+1)(1-\cN/N)   \\
	 &+  \bigg( \sqrt{N} b\left(  \left( N^3 V(N\cdot) * \vert \pn\vert^2   -8\pi \frak{a}_0|\pn|^2 \right)\pn\right)\\
	&\hspace{0.5cm} - \frac{\mathcal{N}+1}{\sqrt N}  b\left( \left( N^3 V(N\cdot) * \vert \pn\vert^2 \right) \pn \right) + \text{h.c.} \bigg)\\
	 &+ \int dx\;     a_x^* (-\Delta_x) a_x +  \int dx\;  V_\text{ext}(x)a_x^{*} a_x  
		\\
		&\hspace{0.5cm} + \int dxdy\;  N^3 V(N(x-y)) \pn^2(y)  \Big(b_x^* b_x - \frac{1}{N} a_x^* a_x \Big)   \\
		&\hspace{0.5cm}+ \int   dxdy\; N^3 V(N(x-y)) \pn(x) \pn(y) \Big( b_x^* b_y - \frac{1}{N} a_x^* a_y \Big)  \\
		&\hspace{0.5cm}+ \frac{1}{2}  \int  dxdy\;  N^3 V(N(x-y)) \pn(y) \pn(x) \Big(b_x^* b_y^*  +  b_xb_y \Big) \\
	&+ \int  dxdy\; N^{ 5/2} V(N(x-y)) \pn(y) \big( b_x^* a^*_y a_x +   a^*_xa_yb_x\big)   \\
	 &+ \frac{1}{2} \int  dxdy\; N^2 V(N(x-y)) a_x^* a_y^* a_y a_x .
	\end{split}
	\end{equation}	
Here, $a_x, a_y^*$, for $x,y \in \bR^3$, denote the usual creation and annihilation operators, satisfying the canonical commutation relations $[a_x, a_y^*] = \delta(x-y)$ and  $[a_x, a_y] = [a_x^*, a_y^*] = 0$, and the operators $ b_x, b^*_y $ denote modified creation and annihilation operators, defined by
		\[ \label{eq:bb-def} b_x = \sqrt{1- \cN/N} \, a_x , \qquad b^*_x = a^*_x \, \sqrt{1- \cN/N}. \]
The operators $b_x, b^*_y$ preserve the number of particles truncation in $\cF_{\perp \pn}^{\leq N}$ and we have  
 		\[
 		[ b_x, b_y^* ] =  ( 1 -  N^{-1}\cN ) \delta (x-y) -  N^{-1} a_y^* a_x, \hspace{0.5cm}[ b_x, b_y ] = [b_x^* , b_y^*] = 0. 
 		\]
Hence, on states with a low number of excitations $\cN\ll N$, the modified field operators satisfy the usual commutation relations up to errors that vanish in the limit $N\to \infty$. 

Extracting the low energy spectrum of $H_N$ by analyzing $\cL_N$ directly seems difficult, because we face a number of serious problems. First, the constant contribution in \eqref{eq:cLNj},
		\[ N \big\langle \pn, \big( -\Delta + V_\text{ext} + \frac{1}{2} \big(N^3 V(N\cdot) * \vert \pn \vert^2\big) \big) \pn \big\rangle   = N\cE_{GP}(\pn) + \mathcal{O}(N),  \]
is off by a quantity of order $\mathcal{O}(N)$ from the correct leading order energy $N\cE_{GP}(\pn)$. Second, the terms that are linear in the creation and annihilation operators,
		\[ \sqrt{N} b\left(  \left( N^3 V(N\cdot) * \vert \pn\vert^2   -8\pi \frak{a}_0|\pn|^2 \right)\pn\right) + \text{h.c.}, \]
are a priori of size $ \mathcal{O}(N^{1/2})$ and, similarly, the simple form bounds
		\[\begin{split} &\pm \frac{1}{2}  \int  dxdy\;  N^3 V(N(x-y)) \pn(y) \pn(x) \big(b_x^* b_y^*  +  b_xb_y \big) \\
		 &\hspace{0.2cm}\leq \frac{1}{2} \int  dxdy\; N^2 V(N(x-y)) a_x^* a_y^* a_y a_x + \frac N2  \| V \|_1\|\pn\|_\infty^2\|\pn\|_2^2 \end{split}\]
indicate that the off-diagonal pairing terms contribute to order $ \mathcal{O}(N)$. The reason for these difficulties is that through the map $U_N$, we expand $H_N$ around the energy of the state $\pn^{\otimes N}$, which neglects the correlations among the particles. The correlations, however, are crucial even for extracting the leading order term $N\cE_{GP}(\pn)$. 

In order to resolve these difficulties, the first step of our analysis consists of renormalizing $\cL_N$ through conjugation by a suitable unitary map that extracts the missing correlation energies of order $\mathcal{O} (N)$. Here, we make use of ideas that were first implemented in the dynamical context \cite{BDS,BS}. To motivate the approach heuristically, let's consider the right hand side in \eqref{eq:cLNj} for the moment as an operator in $\cF^{\leq N} = \bigoplus_{n=0}^N L^2_s(\bR^{3n})$, ignoring the orthogonality constraints in $\cFpn$. Setting 
		\[ \cH_N = \int dx\;     a_x^* (-\Delta_x) a_x +  \int dx\;  V_\text{ext}(x)a_x^{*} a_x   +  \frac{1}{2} \int  dxdy\; N^2 V(N(x-y)) a_x^* a_y^* a_y a_x \]
and $\wt \eta (x;y)=(-N)(1-f)(N(x-y))\pn (x)\pn(y)$, recalling that $f$ denotes the solution of the zero energy scattering equation \eqref{eq:0en}, as well as
		\[  B (\wt \eta)  = \frac12\int dx dy\; \wt \eta(x;y) b^*_xb^*_y - \text{h.c.},	\]
we are going to conjugate the terms on the right hand side in \eqref{eq:cLNj} by the unitary operator exponential $e^{B(\wt \eta)}:\cF^{\leq N}\to \cF^{\leq N}$, using the second order Taylor expansion
		\[  e^{-B(\wt \eta)} O \,e^{B(\wt \eta)} \approx O + [ O, B(\wt \eta) ] +\frac12 \big[[O,B(\wt \eta)], B(\wt \eta)\big] \]	
for observables $O$ in $\cF^{\leq N}$. The main observation that follows from \cite{BDS,BS} is that conjugation by $ e^{ -B(\wt \eta)}(\cdot)e^{ B(\wt \eta)}$ leads to the renormalizations
		\be \label{eq:canc1} \begin{split} 
		& e^{-B(\wt \eta)} \Big(\sqrt{N} b\left(  \left( N^3 V(N\cdot) * \vert \pn\vert^2   -8\pi \frak{a}_0|\pn|^2 \right)\pn\right) + \text{h.c.}\Big)   e^{B(\wt \eta)}  \\
		 &\hspace{0.5cm}+  \int dxdy\;  N^{ 5/2} V(N(x-y)) \pn(y) e^{-B(\wt \eta)}\big( b_x^* a^*_y a_x +   a^*_xa_yb_x\big)  e^{ B(\wt \eta)}\\
		& \approx  \int dxdy\;  N^{ 5/2} V(N(x-y)) \pn(y)  \big( b_x^* a^*_y a_x +   a^*_xa_yb_x\big)  
		\end{split} \ee
and
		\be \label{eq:canc2} \begin{split} 
		& N \big\langle \pn, \big( -\Delta + V_\text{ext} + \frac{1}{2} \big(N^3 V(N\cdot) * \vert \pn \vert^2\big) \big) \pn \big\rangle + e^{ -B(\wt \eta)}\cH_N e^{ B(\wt \eta)}  \\
		 &\hspace{0.5cm}+ \frac{1}{2}  \int  dxdy\;  N^3 V(N(x-y)) \pn(y) \pn(x) e^{-B(\wt \eta)}\Big(b_x^* b_y^*  +  b_xb_y \Big) e^{ B(\wt \eta)} \\
		& \approx  N\cE_{GP}(\pn) +   \cH_N, 
		\end{split} \ee
while the remaining parts on the r.h.s. in \eqref{eq:cLNj} are essentially left unchanged, up to errors of order $\mathcal{O}(1)$ which are well under control. In other words, conjugation by $ e^{ -B(\wt \eta)}(\cdot)e^{ B(\wt \eta)}$ leads to the cancellation of the large linear and quadratic pairing terms in \eqref{eq:cLNj}, and it renormalizes the constant $ \mathcal{O}(N)$ contribution to the correct energy $N\cE_{GP}(\pn)$. 

To make the above heuristics rigorous, we need to make a few technical adjustments. First of all, instead of conjugating with the exponential of $B(\wt \eta)$, we choose a related operator $B=B(\eta)$ with kernel $\eta \in (Q\otimes Q)L^2(\bR^3\times \bR^3)$, for $Q = 1- |\pn \rangle\langle \pn |$. This ensures that $e^B$ is a unitary map from $  \cFpn $ to itself. Moreover, if $\|\eta\|_2$ is sufficiently small, we know from \cite{BS, BBCS3, BBCS4, BSS1} that one has the exact expansion   
		\be \label{eq:defd} e^{-B} b(g) e^{B} = b( \cosh_\eta (g)) + b^*( \sinh_\eta (\overline{g})) + d_\eta (g), \ee
for every $ g \in L^2_{\perp \pn}  (\bR^3)$, where $b(g) = \int_{\bR^3}dx\, \overline{g}(x)b_x $ and where 	
		\begin{equation}\label{eq:defcoshsinheta}
		\begin{split}
		\cosh_\eta = \sum_{n=0}^\infty\frac{ \eta^{(2n)}}{(2n)!}, \hspace{0.5cm} \sinh_\eta = \sum_{n=0}^\infty\frac{ \eta^{(2n+1)} }{(2n+1)!}.
		\end{split}
		\end{equation}
Moreover, the remainder $ d_\eta (g) $ in \eqref{eq:defd} is small on states with a small number $\cN\ll N$ of excitations, in the sense that $ \| d_\eta (g) \xi \| \leq  C N^{-1} \|g\|  \Vert (\mathcal{N}+1)^{3/2} \xi \Vert$ for all $\xi \in \cFpn$. 

To choose a kernel $\eta \in (Q\otimes Q)L^2(\bR^3\times \bR^3)$ with small $L^2(\bR^3\times \bR^3)$ norm that still leads to the cancellations \eqref{eq:canc1} and \eqref{eq:canc2} (up to lower order corrections), one could directly modify $\wt \eta$ from above, for example by imposing a suitable cutoff in position space (such an approach is carried out in \cite{NT}). We use instead previously established tools from the works \cite{BS, BBCS3, BBCS4, BSS1} and consider the Neumann ground state $f_\ell$ solving  
		\[ \Big( -\Delta + \frac{1}{2} V \Big) f_{\ell} = \lambda_{\ell} f_\ell \]
in the ball $B_{N\ell}$ of radius $ N\ell$, fixing $0 < \ell < 1$, and normalized such that $f_\ell (x) = 1$ if $|x| = N \ell$ (notice that $f_\ell$ is radially symmetric). Note that, by scaling, $f_\ell (N.)$ solves 
		\[ \Big( -\Delta + \frac{ 1}{2}N^2 V (N.) \Big) f_\ell (N.) = N^2 \lambda_\ell f_\ell (N.) \]
in the ball $ B_\ell$ of radius $ \ell$. We then set $w_\ell = 1-f_\ell$ which satisfies the pointwise bounds 
		\[ 0\leq w_\ell(x)\leq \frac{C}{|x|+1} \quad\text{ and }\quad |\nabla w_\ell(x)|\leq \frac{C }{|x|^2+1}. \]  
Defining $\eta = (Q\otimes Q ) k \in (Q\otimes Q)L^2(\bR^3\times \bR^3)$ for $ k(x;y) =  -N w_\ell(N(x-y)) \pn(x)\pn(y)$, it is then simple to check that $\| \eta\|_2\leq C \ell^{1/2} $ so that for $\ell >0$ sufficiently small (independently of $N$), we can make use of the expansions \eqref{eq:defd}.

Using the unitary map $e^B$ with $B=B(\eta)$ defined above, we can now analyze the excitation Hamiltonian $\cG_{N} = e^{-B} \cL_N e^{B}$. Before stating its main properties, let us define
 		\[ \begin{split}
		\cK  \!= \! \int dx\; a^*_x( -\Delta_x) a_x,  \, \cV_\text{ext} \!= \! \int dx\, V_\text{ext}(x)a^*_x a_x, \,\cV_N \!= \!\frac12 \int dxdy\; N^2V(N(x-y))a^*_x a^*_y a_y a_x  \end{split} \]
so that $\cH_N = \cK + \cV_\text{ext} + \cV_N$. The following proposition was proved in \cite[Prop 2.5]{BSS2}.
\begin{prop}  \label{prop:GN}
Assume \eqref{eq:asmptsVVext}, let $\cG_{N}=e^{-B} \cL_N e^{B}$ and let $E_N$ denote the ground state energy of $H_N$, defined in \eqref{eq:defHN}. Then, the following holds true: 
\begin{itemize}
\item[a)] We have that $ |E_N - N\cE_{GP}(\pn)| \leq C$ and 
		%\begin{equation}\label{eq:GDelta} 
		\[\cG_N - E_N = \cH_N + \Delta_N, \]%\end{equation}
where the error term $\Delta_N$ is such that for every $\delta >0$, there exists $C > 0$ with
		%\begin{equation}\label{eq:Delta-bd} 
		\[\pm \Delta_N \leq \delta \cH_N + C (\cN + 1). \]
		%\end{equation}
Furthermore, for every $k \in \bN$ there exists a $C > 0$ such that
		\begin{equation}\label{eq:adkN}
		\pm \text{ad}^{\, (k)}_{\, i\cN} (\cG_N) = \pm \text{ad}^{\, (k)}_{\, i\cN} (\Delta_N) = \pm \big[ i \cN , \dots \big[ i \cN , \Delta_N \big] \dots \big] \leq C (\cH_N + 1). \end{equation}

\item[b)] Let $\sigma = \sinh_\eta$ and $ \gamma = \cosh_\eta$ be defined as in \eqref{eq:defcoshsinheta} and let $\kappa_{\cG_N}$ denote the constant
		\begin{equation}\label{eq:defcN}
		\begin{split}
		\kappa_{\cG_N} & = N\big\langle \pn, (-\Delta +V_\text{ext} + \wh V(0)\pn^2/2)\pn\big\rangle- 4\pi \mathfrak{a}_0\|\pn\|_4^4  \\
		&\hspace{0.5cm}  + \tr\big(\sigma (-\Delta+V_\text{ext} -\eps_{GP})\sigma\big) \\
		&\hspace{0.5cm}+ \tr\big( \gamma   \big[N^3 V(N(x-y)) \pn(x)\pn(y)\big]  \sigma ) \\
		&\hspace{0.5cm}  +\tr\big( \sigma \big[   N^3 V(N.) \ast \pn^2  + N^3 V(N(x-y)) \pn(x)\pn(y)\big] \sigma\big)\\
		&\hspace{0.5cm} + \frac12  \int dxdy\, N^2V(N(x-y))| \langle\sigma_x, \gamma_y\rangle|^2. 
		\end{split}
		\end{equation}
Here,  $(N^3V(N.)\ast \pn^2)$ acts as multiplication operator and we identify kernels like $ N^3V(N(x-y))\pn(x)\pn(y)$ with their associated Hilbert Schmidt operators. Let
		%\begin{align*}\label{eq:defPhi}
		%\begin{equation}\label{eq:Phi-def} 
		\[\begin{split}
		\Phi &=  \gamma \big(-\Delta+V_\text{ext}-\eps_{GP}\big)\gamma + \sigma \big(-\Delta+V_\text{ext}-\eps_{GP}\big)\sigma\\
		&\hspace{0.4cm}+ \gamma \big( N^3V(N.)\ast\pn^2 + N^3V(N(x-y))\pn(x)\pn(y)\big)\gamma \\
		&\hspace{0.4cm}+ \sigma\big(N^3V(N.)\ast\pn^2 + N^3V(N(x-y))\pn(x)\pn(y)\big)\sigma\\
		&\hspace{0.4cm}+\Big( \gamma \big(  N^3(Vf_\ell)(N(x-y))\pn(x)\pn(y)\big)\sigma+\emph{h.c.}\Big),
		\end{split}\]%\end{equation} 
and 
		 
		\begin{equation} \label{eq:Gamma-def} 
		\begin{split}
		\Gamma &=  \gamma \big(  N^3(Vf_\ell)(N(x-y))\pn(x)\pn(y)\big)\gamma\\
		&\hspace{0.5cm}+\sigma \!\big( N^3(Vf_\ell)(N(x-y))\pn(x)\pn(y)\big)\!\sigma\\
		&\hspace{0.5cm} +\big[ \sigma \big(-\Delta+V_\text{ext}-\eps_{GP}\big)\gamma +\emph{h.c.}\big]\\
		&\hspace{0.5cm}+ \big[ \sigma \big(N^3V(N.)\ast\pn^2 + N^3V(N(x-y))\pn(x)\pn(y)\big)\gamma + \emph{h.c.}\big].\\
		\end{split}\end{equation} 
With $\Phi$ and $\Gamma$, we define the quadratic Fock space Hamiltonian
		\begin{equation}\label{eq:cQcGN}
		\begin{split}
		\cQ_{\cG_N} = \int dxdy\, \Phi(x;y) b^*_x b_y + \frac12 \int dxdy\,  \Gamma(x;y) \big( b^*_x b^*_y + b_xb_y\big),
		\end{split}
		\end{equation}
and we denote by $\cC_{\cG_N}$ the cubic Fock space operator
		\begin{equation}\label{eq:cCcGN}
		\cC_{\cG_N}=  \int dxdy \, N^{5/2}V(N(x-y)) \pn(y)  b^*_x b^*_y \big(b (\gamma_x) + b^*(\sigma_x)\big) +\emph{h.c.},
		\end{equation}
where $\gamma_x(\cdot) = \gamma(\cdot;x)$ and $\sigma_x(\cdot) = \sigma(\cdot;x)$ for $x\in \bR^3$. Then, $\cG_N$ is equal to
		\begin{equation}\label{eq:propGN}
		\begin{split}
		\mathcal{G}_{N } 
		&=  \kappa_{\cG_N} + \cQ_{\cG_N} + \cC_{\cG_N} +\cV_N + \cE_{\cG_N}
		\end{split}
		\end{equation}
for some self-adjoint operator $ \cE_{\cG_N}$ which is bounded in $\cFpn$ by
		 \begin{equation} \label{eq:cEcGNbnd}
		 \pm \cE_{\cG_N} \leq  CN^{-1/2} (\cH_N +\cN^2+ 1)(\cN+1).
		 \end{equation}
\end{itemize}		
\end{prop}
Proposition \eqref{prop:GN} implements the heuristic renormalizations in \eqref{eq:canc1} and \eqref{eq:canc2} rigorously, noticing that $ \kappa_{\cG_N}= N\cE_{GP}(\pn) +\mathcal{O}(1)$ and $ \| \Gamma \|_{2} \leq C$, uniformly in $N\in \bN$, for $ \kappa_{\cG_N}$ and $\Gamma$ defined in \eqref{eq:defcN} and \eqref{eq:Gamma-def}, respectively (see \cite{BSS2} for the details). Moreover, the error $\cE_{\cG_N}$ in \eqref{eq:propGN} satisfying \eqref{eq:cEcGNbnd}, is negligible on low energy eigenstates. Indeed, this is a direct consequence of the following proposition which was proved in \cite[Theorem 2.6]{BSS2} by combining the main result \eqref{eq:HNlwrbnd} of \cite{BSS1} with the commutator estimates \eqref{eq:adkN}. 
\begin{prop}\label{prop:hpN}
Assume \eqref{eq:asmptsVVext} and let $E_N$ denote the ground state energy of $H_N$, defined in \eqref{eq:defHN}. For $\zeta > 0$, let $\psi_N \in L^2_s (\mathbb{R}^{3N})$ with $\| \psi_N \| = 1$ be such that
		\[ \psi_N = {\bf 1}_{(-\infty ; E_N + \zeta]} (H_N) \psi_N\]
and let $\xi_N = e^{-B} U_N \psi_N\in \cFpn$ denote the renormalized excitation vector related to $\psi_N$. Then, for every $j\in\mathbb N$ there is a constant $C > 0$ such that 
\begin{align*}%\label{eq:hpN} 
\langle \xi_N, (\cH_N+1) (\cN +1)^j  \xi_N \rangle \leq C (1 + \zeta^{j+1}). \end{align*}
\end{prop}
 
With Bogoliubov's arguments \cite{Bog} in mind, we may be tempted to neglect the cubic and quartic contributions $\cC_{\cG_N}$ and $\cV_N$ on the r.h.s. in \eqref{eq:cCcGN} and diagonalize the remaining quadratic operator $\cQ_{\cG_N}$ in \eqref{eq:cQcGN} to determine the spectrum of $\cG_N$ up to order $\mathcal{O}(1)$. Proceeding this way does, however, not yield the right spectrum, because the terms $\cC_{\cG_N}$ and $\cV_N$ still contain important energy contributions of order $\mathcal{O}(1)$. This has been a key observation for the analysis in \cite{BBCS3} (see also \cite{YY} and \cite{NRS1, NRS2} related to this point) so that, in order to conclude Theorem \ref{thm:main}, we first need to extract the missing $\mathcal{O}(1)$ energies. 

To this end, we proceed similarly as in the first step, now having the goal to renormalize the cubic contribution $\cC_{\cG_N}$ in \eqref{eq:cCcGN}. To understand the main idea on a heuristic level, consider for simplicity first the cubic operator
		\be \label{eq:cubicmain} \int dxdy \, N^{5/2}V(N(x-y)) \pn(y)  b^*_x b^*_y b_x  +\text{h.c.} \ee
that is obtained from $\cC_{\cG_N}$ by replacing $\gamma_x \approx \delta_x$ and $\sigma_x\approx 0$ (to leading order in $\eta$). By analogy to the quadratic renormalization through $e^B$, a straightforward computation involving the scattering equation \eqref{eq:0en} implies that 
 		\[\begin{split}
		&\int dxdy \,  (-N (1-f)(N(x-y)) \pn(y)  [ \cK_N +\cV_N, b^*_x b^*_y b_x  - \text{h.c.} ] \\
		&\hspace{0.5cm}\approx -\int dxdy \, N^{5/2}V(N(x-y)) \pn(y)  b^*_x b^*_y b_x  +\text{h.c.}, 
		\end{split}\]
up to errors of order $\mathcal{O}(1)$. What this indicates is that we can cancel the cubic term \eqref{eq:cubicmain} by conjugating $\cG_N$ through a unitary operator exponential that is now cubic in the (modified) creation and annihilation operators, having the form
		\be\label{eq:Arough}\wt A  =  \frac{1}{\sqrt{N}} \int dx dy \, (-N (1-f)(N(x-y)) \pn(y)   {b}_x^* {b}_y^*  b_x   - \text{h.c.}\ee 

From the analysis of the quadratic renormalization and the fact that our goal is to cancel the full cubic term $\cC_{\cG_N}$ in $\cG_N$ (and not only the term in \eqref{eq:cubicmain}), it is clear that we need to make a few technical adjustments to $\wt A$ in \eqref{eq:Arough}. First of all, to stay consistent with the first renormalization step, instead of working with the correlation factor $ (- N(1-f))(N.) $, we use the function $(-N w_\ell)(N.)  $ on which we impose additionally an $N$-dependent low momentum cutoff. More precisely, for some small $\exph > 0$, we set $\chi_H (p) = \chi (|p| > N^\exph)$, denote by $\check\chi_H$ its inverse Fourier transform and define the kernel
		\begin{equation}\label{eq:def-wtk} \wt{k}_H (x;y) = \left( - N w_\ell (N.) * \check{\chi}_H \right) (x-y) \ph_0 (y). \end{equation} 
Through the cutoff $\check \chi_H$, it turns out that it is enough to compute only a few commutators in the expansion of the cubic conjugation, because most of the error terms become small (in orders of $N$).  Second, to cancel not only the cubic term in \eqref{eq:cubicmain}, but the full term $\cC_{\cG_N}$ in \eqref{eq:cCcGN}, we need to replace the $b_x$-field in \eqref{eq:Arough} by $  b (\gamma_x) + b^*(\sigma_x) $. Also here, it turns out to simplify the analysis if we impose high momentum cutoffs on the kernels $\gamma_x$ and $\sigma_x$. To this end, we choose $0< \exps < \exph$, we set $g_L (p) = e^{-p^2/ N^{2\tau}}$ and we denote by $  \check{g}_L$ its inverse Fourier transform. With the notation  
		\[ \sigma_{L} = \sigma *_2 \check{g}_L , \qquad \gamma_{L} = \gamma *_2 \check{g}_L, \]
where $*_2$ denotes convolution in the second variable, we then define the cubic operator 
		\[  A =  \frac{1}{\sqrt{N}} \int dx dy \, \wt{k}_H (x;y) \wt{b}_x^* \wt{b}_y^* \left[ \wt{b} (\gamma_{L,x})  + \wt{b}^* (\sigma_{L,x})  \right] - \text{h.c.}
		\] 
where $ \wt{b}_x = b (Q_x) = \int dz \, Q (x;z) b_z $, with $Q = 1 - |\ph_0 \rangle \langle \ph_0|$. Notice that the use of the operators $ \wt b_x, \wt b^*_y$ ensures that $A$ is a unitary map from $\cF_{\perp \ph_0}^{\leq N}$ to itself. 

The following proposition summarizes the main properties of the renormalized excitation Hamiltonian $\cJ_N = e^{-A} \cG_N e^{A}$. For its proof, see  \cite[Prop. 2.8]{BSS2}. 

\begin{prop} \label{prop:JN}
	Assume \eqref{eq:asmptsVVext} and let $0<6 \exps\leq \exph\leq \frac{1}{2}$. Then
	\begin{equation}\label{eq:JNdec}
	\mathcal{J}_N = \kappa_{\mathcal{J}_N} + \mathcal{Q}_{\mathcal{J}_N} + \mathcal{V}_N + \mathcal{E}_{\mathcal{J}_N},
	\end{equation}
	where 
\be \label{eq:kJN}\begin{split}
	\kappa_{\mathcal{J}_N} 
	&= \kappa_{\cG_N}  -\emph{tr}(\sigma N^3 (Vw_\ell)(N.) * \pn^2 \sigma) - \emph{tr}(\sigma N^3 (Vw_\ell)(N(x-y)) \pn(x)\pn(y) \sigma)
	\end{split}\ee
	with $\kappa_{\cG_N}$ defined in \eqref{eq:defcN} and where the quadratic operator $ \cQ_{\mathcal{J}_N}$ is given by
	%\begin{equation}\label{eq:cQcJN}
	\[\begin{split}
	\cQ_{\mathcal{J}_N} = \int dxdy\, \widetilde{\Phi}(x;y) b^*_x b_y + \frac12 \int dxdy\, \widetilde{ \Gamma}(x;y) \big( b^*_x b^*_y + b_xb_y\big)
	\end{split}
	\]%\end{equation}
	for
	\begin{equation}\label{eq:deftildePhi}
	\begin{split}
	\widetilde{\Phi} &=  \gamma \big(-\Delta+V_\text{ext}-\eps_{GP}\big)\gamma + \sigma \big(-\Delta+V_\text{ext}-\eps_{GP}\big)\sigma\\
	&\hspace{0.4cm}+ \gamma \big( 8\pi\frak{a}_0\pn^2 + N^3(Vf_\ell)(N(x-y))\pn(x)\pn(y)\big)\gamma \\
	&\hspace{0.4cm}+ \sigma\big(8\pi\frak{a}_0\pn^2 + N^3(Vf_\ell)(N(x-y))\pn(x)\pn(y)\big)\sigma\\
	&\hspace{0.4cm}+\Big( \gamma \big(  N^3(Vf_\ell)(N(x-y))\pn(x)\pn(y)\big)\sigma+\emph{h.c.}\Big),
	\end{split}
	\end{equation}
	and 
	\begin{equation}\label{eq:deftildeGamma}
	\begin{split}
	\widetilde{\Gamma} &=  \gamma \big(  N^3(Vf_\ell)(N(x-y))\pn(x)\pn(y)\big)\gamma\\
	&\hspace{0.5cm}+\sigma \big( N^3(Vf_\ell)(N(x-y))\pn(x)\pn(y)\big) \sigma\\
	&\hspace{0.5cm} +\big[ \sigma \big(-\Delta+V_\text{ext}-\eps_{GP}\big)\gamma +\emph{h.c.}\big]\\
	&\hspace{0.5cm}+ \big[ \sigma \big(8\pi\frak{a}_0\pn^2 + N^3(Vf_\ell)(N(x-y))\pn(x)\pn(y)\big)\gamma + \emph{h.c.}\big].\\
	\end{split}
	\end{equation}
	Moreover, the self-adjoint operator $\mathcal{E}_{\mathcal{J}_N}$ is bounded by 
		\begin{equation} \label{eq:errorJN}
	\pm \mathcal{E}_{\mathcal{J}_N} \leq  C N^{- \min \{\frac{\exps}{2}, \frac{1}{2}-\exph\}}  (\mathcal{H}_N+1)(\mathcal{N}+1)^3.
	\end{equation}
\end{prop}
Compared to Prop. \ref{prop:GN}, notice that the decomposition \eqref{eq:JNdec} of $\cJ_N$ does not contain a cubic contribution anymore. In other words, conjugating $\cG_N$ by the cubic exponential $e^{-A}(\cdot)e^A$ establishes rigorously the cubic renormalization outlined after \eqref{eq:cubicmain}. To control the error term $\cE_{\cJ_N}$ in \eqref{eq:JNdec}, we make use of the error bound \eqref{eq:errorJN} and the following analogue of Prop. \ref{prop:hpN} (see \cite[Prop. 2.9]{BSS2} for the proof).

\begin{prop}\label{prop:hpNcubic}
Assume \eqref{eq:asmptsVVext} and let $E_N$ denote the ground state energy of $H_N$. For some $\zeta > 0$, let $\psi_N \in L^2_s (\mathbb{R}^{3N})$ with $\| \psi_N \| = 1$ be such that
	 \[ \psi_N = {\bf 1}_{(-\infty ; E_N + \zeta]} (H_N) \psi_N. \]
Let $\xi_N =e^{-A} e^{-B} U_N \psi_N\in \cFpn$ be the renormalized excitation vector related to $\psi_N$. Then, for any $j \in\mathbb N$ there exists some $C > 0$ such that 
	\[ \langle \xi_N, (\cN +1)^j (\cH_N+1) \xi_N \rangle \leq C (1 + \zeta^{j+3}). \]
\end{prop}
	
%%%%%%%%%%%%%%%%%%%%%%%%%%%%%%%%%%%%%%%%%%%%%%%%%%%%%%%%%%%%
%%%%%%%%%%%%%%%%%%%%%%%%%%%%%%%%%%%%%%%%%%%%%%%%%%%%%%%%%%%%
%%%%%%%%%%%%%%%%%%%%%%%%%%%%%%%%%%%%%%%%%%%%%%%%%%%%%%%%%%%%
\section{Diagonalization and Proof of Theorem \ref{thm:main}}  

Using the decomposition \eqref{eq:JNdec} in Prop. \ref{prop:JN}, we can now determine the spectrum $\sigma(\cJ_N)$ of $\cJ_N= e^{-A}e^{-B}U_NH_NU_N^*e^Be^A$ (and hence of $H_N$, by unitary equivalence), up to errors that vanish as $N\to\infty$. We choose the momentum cutoff parameters (introduced around \eqref{eq:def-wtk}) $\eps = 6/13$ and $\tau =\eps/6$, so that we can apply Propositions \ref{prop:GN}, \ref{prop:hpN}, \ref{prop:JN} and \ref{prop:hpNcubic}. 

To obtain upper and lower bounds on $\sigma(\cJ_N)$, we apply the min-max principle and compare the eigenvalues of $\cJ_N$ with those of the quadratic Fock space Hamiltonian
		%\begin{equation} \label{eq:deftildeQ}
		\[\widetilde{\mathcal{Q}}_{\mathcal{J}_N} = \kappa_{\cJ_N} +  \int dx dy \, \widetilde{\Phi}(x,y) a_x^* a_y + \frac{1}{2}\int dx dy \ \widetilde{ \Gamma}(x,y) (a_x^* a_y^* + \text{h.c.}),
		\]%\end{equation}
defined in the Fock space $\cF_{\bot \pn} = \bigoplus_{n=0}^\infty L^2_{\bot \pn}(\bR^3)^{\otimes_s n}$ (without a restriction on the number of particles). Here, $\kappa_{\cJ_N}$ denotes the constant from \eqref{eq:kJN}, and the kernels of $\wt \Phi$ and $ \wt \Gamma$ were defined in \eqref{eq:deftildePhi} and \eqref{eq:deftildeGamma}, respectively.

To see that it is enough to analyze $\widetilde{\mathcal{Q}}_{\mathcal{J}_N} $, notice first that, to get a lower bound on the min-max values of $\cJ_N$, we can drop the non-negative potential energy $\cV_N$ in \eqref{eq:JNdec} and we can control the error $\cE_{\cJ_N}$ in \eqref{eq:JNdec} through Prop. \ref{prop:hpNcubic}. Since $\cFpn\subset \cF_{\bot \pn}$, we obtain a lower bound on the spectrum of $\cJ_N$ by computing the spectrum of $\widetilde{\mathcal{Q}}_{\mathcal{J}_N}$ in $\cF_{\bot \pn}$. To obtain upper bounds on the min-max values of $\cJ_N$, on the other hand, we construct suitable subspaces in $\cFpn$, built from the eigenvectors of $\widetilde{\mathcal{Q}}_{\mathcal{J}_N}$ truncated to $\cFpn$. It turns out that the potential energy $\cV_N$ is negligible on such eigenspaces and that the upper and lower bounds on the min-max values of $\cJ_N$ coincide, up to errors vanishing in the limit $N\to \infty$ (see \cite[Section 3]{BSS2} for the details).

Hence, let us focus on $\widetilde{\mathcal{Q}}_{\mathcal{J}_N}$ and let us recall how to determine its spectrum. As a quadratic Fock space Hamiltonian, $\widetilde{\mathcal{Q}}_{\mathcal{J}_N}$ is exactly diagonalizable. To diagonalize it, we follow $\cite{GS}$ and recall the definitions of $H_\text{GP}$ and $E$ in \eqref{eq:defHGPE} and \eqref{eq:defHGPE2}, respectively. We then identify the operators $H_{GP}, E, \wt \Phi, \wt \Gamma$ with operators mapping $L^2_{\bot \pn}(\bR^3)$ back into itself (so that, in particular, $H_\text{GP}$ and $E$ are invertible in $L^2_{\bot \pn}(\bR^3)$). Moreover, we set
		%\be \label{eq:deftildeDE} \begin{split}
		\[\begin{split}\wt{D} &= \widetilde{\Phi} - \widetilde{ \Gamma}  \hspace{0.5cm} \text{and} \hspace{0.5cm}\wt{E}  = \big( \wt{D}^{1/2} ( \wt{D} + 2\wt\Gamma) \wt{D}^{1/2}\big)^{1/2} 
		\end{split} \]
as well as $A = {\wt{D}}^{1/2} \wt{E}^{-1/2}$ and $ \alpha = \log\big( \vert A^* \vert\big)$. Finally, Writing  $ A =W \vert A \vert $ for some partial isometry in $L_{\perp \pn}^2(\mathbb{R}^3)$, by the polar decomposition, we construct a standard Bogoliubov transformation $\cU$ that diagonalizes $\wt{\cQ}_{\cJ_N}$. Indeed, denoting by $(\varphi_j)_{j\in \mathbb{N}}$ the eigenbasis of $\wt E$ and setting $a^\sharp_j  = a^\sharp (\varphi_j)$ for $\sharp\in \{\cdot, *\}$ as well as $\alpha_{ij} = \langle \varphi_i, \alpha\, \varphi_j\rangle$, we set
		\[ \cW = \Gamma(W), \hspace{0.5cm} X  = \frac12 \sum_{i, j =1}^\infty \alpha_{ij} a^*_i a^*_j -\text{h.c.}, \hspace{0.5cm} \cU = e^{X} \cW.   \]
Here, $ \Gamma(W)$ denotes the second quantization of the partial isometry $W$, acting in the $n$-particle sector of $\cF_{\bot \pn}$ as $W^{\otimes n}$. With $\cU$ defined above, one verifies that
		\be\label{eq:diag}\begin{split}
		\cU^* \wt{\cQ}_{\cJ_N} \cU & =  \kappa_{\cJ_N} +  \frac12 \tr_{\bot \pn} \bigg( \frac{1}{2}\big(\wt{D}^{1/2} \wt{E} \wt{D}^{-1/2} + \wt{D}^{-1/2} \wt{E} \wt{D}^{1/2}\big) - \wt D - \wt \Gamma\bigg) + d\Gamma(\wt E),
		\end{split}\ee
where $ \tr_{\bot \pn} $ denotes the trace in $L^2_{\bot \pn}(\bR^3)$ and where $d\Gamma(\wt E)$ denotes the second quantization of $\wt E$, acting as $\sum_{i=1}^n \wt E_{x_i}$ in the $n$-particle sector of $\cF_{\bot \pn}$. 

Now, using that the eigenvalues of $d\Gamma( \wt E )$ are equal to those of $ d\Gamma(E)$ (with $ E$ defined in \eqref{eq:defHGPE2}), up to errors that vanish as $N\to \infty$, we conclude \eqref{eq:excitHN-EN}. Theorem \ref{thm:main} then follows by showing that the constant on the right hand side in \eqref{eq:diag}, that is
		\[\kappa_{\cJ_N} +  \frac12 \tr_{\bot \pn} \bigg( \frac{1}{2}\big(\wt{D}^{1/2} \wt{E} \wt{D}^{-1/2} + \wt{D}^{-1/2} \wt{E} \wt{D}^{1/2}\big) - \wt D - \wt \Gamma\bigg), \]
is equal to the constant on the right hand side in \eqref{eq:gsenergy}, up to errors that vanish as $N\to\infty$. For the details of this computation, see \cite[Section 3]{BSS2}.

%%%%%%%%%%%%%%%%%%%%%%%%%%%%%%%%%%%%%%%%%%%%%%%%%%%%%%%%%%%%
%%%%%%%%%%%%%%%%%%%%%%%%%%%%%%%%%%%%%%%%%%%%%%%%%%%%%%%%%%%%
%%%%%%%%%%%%%%%%%%%%%%%%%%%%%%%%%%%%%%%%%%%%%%%%%%%%%%%%%%%%

\end{document}